\documentclass[12pt]{iopart}
\usepackage{bm}
\usepackage{graphicx}

\begin{document}

\title{Transients from initial conditions based on
Lagrangian perturbation theory \\
in $N$-body simulations II: the effect of the transverse mode}

\author{Takayuki Tatekawa$^{\dag,\ddag}$}

\address{\dag Center for Information initiative,
3-9-1 Bunkyo, Fukui, Fukui 910-8709, JAPAN}
\address{\ddag Research Institute for Science and Engineering,
Waseda University, 3-4-1 Okubo, Shinjuku,
Tokyo 169-8555, JAPAN}

\begin{abstract}

We study the initial conditions for cosmological $N$-body simulations
for precision cosmology. In general, Zel'dovich approximation
has been applied for the initial conditions of $N$-body
simulations for a long time. These initial conditions provide
incorrect higher-order growth. These error caused by
setting up the initial conditions by perturbation theory is called
transients. We investigated the impact of transient on non-Gaussianity
of density field by performing cosmological $N$-body simulations
with initial conditions based on first-, second-, and third-order
Lagrangian perturbation theory in previous paper. In this paper,
we evaluates the effect of the transverse mode in the third-order
Lagrangian perturbation theory for several statistical quantities
such as power spectrum and non-Gaussianty.
Then we clarified that the effect of the transverse mode in
the third-order Lagrangian perturbation theory is quite small.

\end{abstract}

\maketitle


\section{Introduction}
 \label{sec:Intro}
 
The structure formation in the Universe is one of the most
important problem in modern cosmology. The primordial
density fluctuation has clustered by its self-gravitating instability.
On the other hand, the cosmic expansion avoids the formation
of the structure. The structure has been formed on the balance
of self-contraction and cosmic expansion
\cite{Peebles,Padmanabhan1993,Sahni1995,Peacock,Liddle,Coles,Bernardeau2002,Jones2004,Weinberg2008}. 
Therefore when
we analyze the evolution of the structure, we would clarify
the evolution of the Universe.

The cosmic expansion is affected by contents of the Universe.
Recently, the acceleration of the cosmic expansion is discovered
\cite{Riess1998,Schmidt1998,Perlmutter1999}.
Although the acceleration of the cosmic expansion is caused by
the cosmological constant, the existence of the cosmological constant 
causes several problems
\cite{Weinberg1989}.
Instead of the cosmological constant,
many dark energy models or modified gravitational theory
has been proposed
\cite{Peebles2003,Copeland2006}.
The evolution of the large-scale structure 
would give the restriction to these models.

For the evolution of the large-scale structure in the Universe,
$N$-body simulations have been carried out for long time~\cite{P3M}.
In the simulations, matter distribution is represented with mass points 
and the interaction between mass points is calculated numerically. 
$N$-body simulation can describe strongly nonlinear structure.

However, for N-body simulations, there is a problem about how to
set up the initial conditions.
Even though the naive expectation is that it is better to start simulations as early
as possible, like recombination era, it is well known that this is extremely difficult
numerically. When we start $N$-body simulation at an early era,
we promote the evolution of the density fluctuation with
Lagrangian perturbation theory (LPT) until quasi-nonlinear stage.
(For reviews of LPT, see for example~\cite{Tatekawa04R,Tatekawa2010})
Then we connect LPT and cosmological $N$-body simulations.
In many case, the Zel'dovich approximation (ZA)~\cite{ZA}, i.e., the first-order
LPT (1LPT) has been applied for the initial conditions of $N$-body
simulations for a long time. 

Recently, Crocce \textit{et al.}~\cite{Crocce2006} proposed the improvement
for the initial conditions. Basically, they applied the second-order
LPT (2LPT). With these initial conditions, they calculate the statistical quantities
such as non-Gaussianity and show the effects of transients
related with 2LPT initial conditions decrease much faster than the ones
related with ZA initial conditions, In other words, even if the difference of
the initial conditions between ZA and 2LPT is quite small,
the difference appears at late time manifestly. 
The transients with 2LPT initial conditions are less harmful
than ones with ZA initial conditions.

Here we have one problem. Is just taking 2LPT into consideration
enough for the initial conditions? Tatekawa and Mizuno~\cite{Tatekawa2007}
considered the transients with ZA, 2LPT, and 3LPT initial conditions
(hereafter, Paper I).
They analyzed the non-Gaussianity of the density field at $0 < z < 5$.
The difference of the non-Gaussianity by the difference between 
2LPT and 3LPT initial conditions was $1$ \% or less. When we
requires percent order accuracy, we should consider 2LPT
initial conditions. Furthermore, when we requires sub-percent order accuracy,
we should consider 3LPT initial conditions. 

In Paper I, only a longitudinal mode in 3LPT has been considered.
Even if we consider only longitudinal mode in ZA, a transverse mode
in 3LPT appears. Although this transverse mode does not have meaning 
of a vorticity by Kelvin's circulation theorem, 3LPT is insufficient 
if the transverse mode is ignored. 

In this paper, we consider both the longitudinal mode and the
transverse mode in 3LPT for the initial conditions for the
cosmological $N$-body simulations. In addition to
the non-Gaussianity of the density field, we analyze
the evolution of the power spectrum.
The power spectrum is one of the most important
statistical quantities in modern cosmology.

This paper is organized as follows. In section~\ref{sec:NonL},
we present Lagrangian perturbative equations valid up to
the third-order. Here we derive both the longitudinal mode
and the transverse mode in 3LPT. The comparison of transient state
is shown in section~\ref{sec:Transient}. First, we briefly explain
cosmological $N$-body simulation. Then we compare 
the power spectrum and the non-Gaussianity of the density field
between with the initial conditions
based on 1LPT, 2LPT, and 3LPT. Section~\ref{sec:Summary}
is devoted to conclusions.


\section{Lagrangian perturbations} 
\label{sec:NonL}

In this section, we briefly summarize evolution equations
of Lagrangian perturbation theory  (LPT)
valid up to the third-order.
In LPT, the motion of
cosmic fluid is determined in Newtonian cosmology.
The approach is based on the continuous equation,
Euler's equation, and Poisson's equation.
The cosmic expansion is given by the Friedmann equations.
The detail of derivation for evolution equations 
were described in Paper I~\cite{Tatekawa2007}.

In this paper, we consider the Lagrangian perturbation
in which solutions for cosmic fluid 
are already derived by several people
\cite{ZA,Buchert1992,Barrow1993,Bouchet1992,Buchert1993,Buchert1994,Bouchet1995,Catelan1995}.
For this purpose, it is necessary to define the comoving Lagrangian
coordinates $\bm{q}$ in terms of the comoving Eulerian coordinates
$\bm{x}$ as:
\begin{equation} \label{eqn:x=q+s}
\bm{x} = \bm{q} + \bm{s} (\bm{q},t) ,
\end{equation}
where $\bm{s}$ is the displacement vector which denotes the deviation from
the homogeneous distribution.

It is worth noting that it is not 
the density contrast $\delta$ but 
the displacement vector $\bm{s}$ that is regarded as
a perturbative quantity in LPT.
In the Lagrangian coordinates, from the continuous equation,
we can express the density contrast exactly as
\begin{equation} 
\label{eqn:L-exactrho}
\delta = J^{-1}-1\,,
\end{equation}
where $J$ is the determinant of the Jacobian of the mapping
between $\bm{q}$ and $\bm{x}$: 
\begin{equation} \label{eqn:Jacobian}
J = 1+ \nabla \cdot \bm{s} + \frac{1}{2} \left ( \left ( \nabla \cdot \bm {s} \right )^2
 - s_{i,j} s_{i,j} \right ) + \det \left ( s_{i,j} \right ) \,,
\end{equation}
where $\nabla$ means the Lagrangian spatial derivative.

From the physical property, 
$\bm{s}$ can be decomposed
to the longitudinal and the transverse modes:
\begin{eqnarray}
&&\bm{s} = \nabla S + \bm{s}^T 
 = g(t) \nabla \psi(\bm{q}) + g^T(t) \bm{\zeta}^T (\bm{q}) \,, \\
&&\nabla \cdot \bm{\zeta}^T = 0 \,,
\end{eqnarray}

The first-order perturbation is known as Zel'dovich perturbation
~\cite{ZA}.
\begin{equation}
\ddot{g}_1 + 2 \frac{\dot{a}}{a} \dot{g}_1
 - 4 \pi G \rho_b g_1 = 0 \label{eqn:1st-L-t-eq} \,.
\end{equation}
Hereafter, we consider only the growing solution.
The spatial part $\psi^{(1)} (\bm{q})$ is given by the initial condition.
The transverse mode does not have growing solutions~\cite{Buchert1992,Barrow1993}.
We ignore the transverse mode in the first-order perturbation.

The second- and third-order perturbative equations are as follows
~\cite{Bouchet1992,Buchert1993,Buchert1994,Bouchet1995,Catelan1995,Sasaki1998}.
\begin{eqnarray}
\ddot{g}_2 + 2 \frac{\dot{a}}{a} \dot{g}_2
 - 4 \pi G \rho_b g_2 &=& - 4 \pi G \rho_b g_1^2 \,,
 \label{eqn:2nd-L-t-eq} \\
\ddot{g}_{3a} + 2 \frac{\dot{a}}{a} \dot{g}_{3a}
 - 4 \pi G \rho_b g_{3a} &=& - 8 \pi G \rho_b g_1 (g_2 - g_1^2 ) ,
 \label{eqn:3rd-L-t-eqA} \\
\ddot{g}_{3b} + 2 \frac{\dot{a}}{a} \dot{g}_{3b}
 - 4 \pi G \rho_b g_{3b} &=& - 8 \pi G \rho_b g_1^3 ,
 \label{eqn:3rd-L-t-eqB}
\end{eqnarray}
\begin{eqnarray}
\psi_{,ii}^{(2)} &=& \frac{1}{2} \left \{
 \psi_{,ii}^{(1)} \psi_{,jj}^{(1)} - \psi_{,ij}^{(1)}
 \psi_{,ij}^{(1)} \right \} \,, \label{eqn:2nd-L-spa-eq} \\
\psi_{,ii}^{(3a)} &=& \frac{1}{2} \left \{
 \psi_{,ii}^{(1)} \psi_{,jj}^{(2)} - \psi_{,ij}^{(1)}
 \psi_{,ij}^{(2)} \right \} , \label{eqn:3rd-L-spa-eqA} \\
\psi_{,ii}^{(3b)} &=& \det \left ( \psi_{,ij}^{(1)} \right ) \nonumber \\
 &=& \frac{1}{6} \psi_{,ii}^{(1)} \psi_{,jj}^{(1)} \psi_{,kk}^{(1)}
 - \frac{1}{2} \psi_{,ii}^{(1)} \psi_{,jk}^{(1)} \psi_{,jk}^{(1)}
 + \frac{1}{3} \psi_{,ij}^{(1)} \psi_{,jk}^{(1)} \psi_{,ki}^{(1)} .
 \label{eqn:3rd-L-spa-eqB}
\end{eqnarray}
Even if we consider only the longitudinal mode in the first-order perturbation,
a third-order perturbation appears in the transverse mode.
\begin{eqnarray}
\ddot{g}_{3T} + 2 \frac{\dot{a}}{a}
\dot{g}_{3T} &=& 4 \pi G \rho_b g_1^3 , \\
- \nabla^2 \zeta_i^{(3)} &=& \left ( \psi_{,il}^{(1)} \psi_{,kl}^{(2)}
 - \psi_{,kl}^{(1)} \psi_{,il}^{(2)} \right )_{,k} .
\end{eqnarray}
Because of Kelvin's circulation theorem, the growing mode of the transverse mode
does not mean evolution of vorticity.

\section{Comparison of transient state} \label{sec:Transient}
\subsection{Numerical Calculations} \label{sec:NumRes}

In this section, we calculate the statistical quantities
introduced in the previous section in the
$\Lambda$CDM model based on $N$-body simulations. 
For setting up the initial conditions, 
we use COSMICS code~\cite{COSMICS}
which generates primordial Gaussian density field
usually based on the ZA.
We consider the case with those based on 2LPT and 3LPT, too.
COSMICS package consists of 4 applications. GRAFIC generated
Gaussian random density fields (density, velocity, and
particle displacements) on a lattice. Both the velocity
and the displacements are related to each other.

GRAFIC automatically selects the output redshift
by the maximum density fluctuation on a grid 
$\delta_{max}$ for a given set
of cosmological parameters. 
In order to obtain the initial redshift, we adopt the
following cosmological parameters
at the present time ($z=0$) which are given
by WMAP 7-year result~\cite{WMAP}:
\begin{eqnarray}
\Omega_m &=& 0.275 \,, \\
\Omega_{\Lambda} &=& 0.725 \,, \\
H_0 &=& 70.2~ \mbox{[km/s/Mpc]} \,, \\
\sigma_8 &=& 0.816 \,, \\
n &=& 0.968 \,.
\end{eqnarray}
We set the input maximum density
$\delta_{\mbox{ini}}=0.3$. The initial redshift
$\bar{z} \simeq 100$.
The initial redshift is set by
the input maximum density fluctuation. Because we
set random Gaussian fluctuation, the initial redshift
is not fixed.

From the initial conditions, we
follow the evolution of the particles based on
N-body simulation.
The initial conditions are given by
ZA (1LPT), 2LPT, 3LPT (longitudinal mode only) and
3LPT (both longitudinal mode and transverse mode).
Hereafter we write 3LPT (longitudinal mode only)
by '3LPT L' .

The numerical algorithm is applied 
by particle-particle particle-mesh
($P^3M$) method~\cite{P3M} which was developed by
Gelb and Bertschinger. The numerical code we use
is written by Bertschinger.
For N-body simulations, we set the parameters as follows:
\begin{eqnarray*}
\mbox{Number of particles} &:& N=512^3 \,, \\
\mbox{Box size} &:& L=512 h^{-1} \mbox{[Mpc]}
 ~~(\mbox{at}~z=0)  \,, \\
\mbox{Softening length} &:& \varepsilon = 50 h^{-1} \mbox{[kpc]}
 ~~(\mbox{at}~z=0)  \,.
\end{eqnarray*}

For the simulations, we use 10 samples for an initial
condition. After the calculations,
in order to avoid the divergence of the density
fluctuation in the limit of large $k$, however,
just for a technical reason,
it is necessary to consider the density field 
$\rho_m(\bm{x};R)$
at the position $\bm{x}$ smoothed over the scale $R$,
which is related to the unsmoothed density field
$\rho_m(\bm{x})$ as
\begin{eqnarray}
\rho_m(\bm{x};R) &=& \int d^3 \bm{y} W(|\bm{x}-\bm{y}|;R)
 \rho_m(\bm{y}) \,,
\end{eqnarray}
where we use the top-hat spherical window function
by Eq.~(\ref{tophat_window}).
\begin{equation}
\tilde{W} = \frac{3(\sin x - x \cos x)}{x^3}\,,
\label{tophat_window}
\end{equation}
Throughout this paper, we choose the smoothing scale
$R=2 h^{-1} \mbox{[Mpc]} ~~(\mbox{at}~z=0)$.

Crocce {\it et al.}~\cite{Crocce2006} analyzed
the non-Gaussianity of the density field with 
both the skewness and the kurtosis.
They changed the smoothing scale $R$ and compared 
the non-Gaussianity with two different initial conditions
which are based on the ZA and 2LPT, respectively.
In Paper I~\cite{Tatekawa2007}, we analyzed
the non-Gaussianity with three different initial
conditions which are based on the ZA, 2LPT, and 3LPT.
From our results, when we set the initial conditions
based on 2LPT at $z \sim 30$, the effects of
the transients on both the skewness and the kurtosis
from 2LPT initial conditions become negligible until
$z \sim 3$. 

In Paper I, we ignored the transverse mode
in 3LPT. As we showed in Sec.\ref{sec:NonL}
even if we consider only longitudinal mode
in ZA, the transverse mode in 3LPT appears.
In this paper,
we consider the effect of the transverse mode
in 3LPT for the non-Gaussianity in the density field.

\subsection{Power spectrum} \label{sec:Spectrum}

First, we analyze the power spectrum in the density
field. The power spectrum is widely applied for 
analyses about nonlinear evolution of the cosmic structure. 
Recently, Baryon Acoustic Oscillation (BAO) has been
discovered~\cite{Eisenstein2005}. The power spectrum
is one of the most powerful tools for the analysis of BAO. 
Because this analysis requires high accurate simulations,
it is meaningful to compare the power spectrum between 
the initial conditions based on ZA and higher-order
perturbations.

\begin{figure}[tb]
\centerline{
\includegraphics[height=14cm]{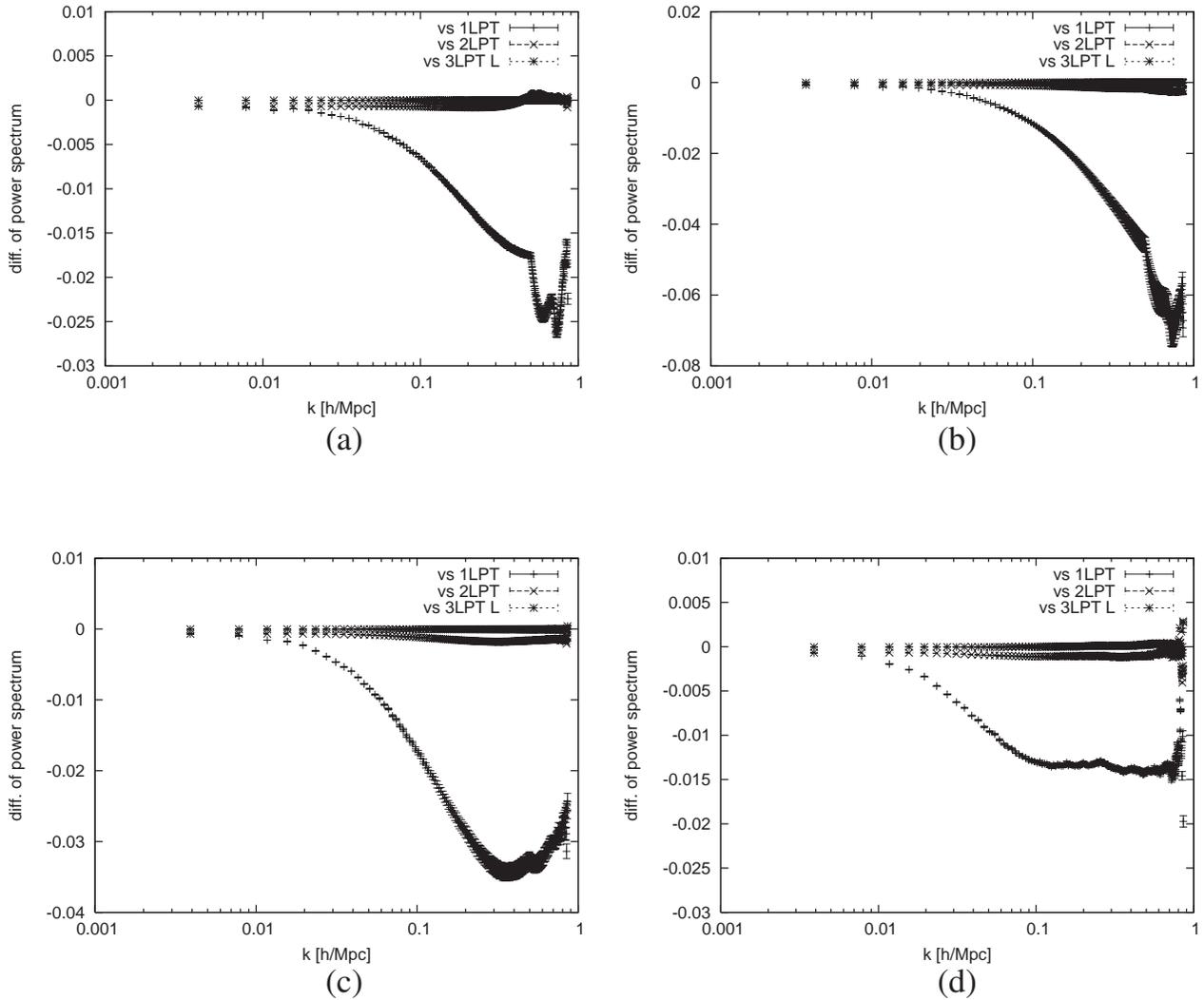}
}
\caption{The difference between the power spectrum 
of the density distribution with the initial conditions based on 
3LPT and other perturbations:
(a) $z=10$, (b) $z=5$, (c) $z=2$,
(d) $z=0$. The error bars are given by the difference between samples.
}
\label{fig:Pk-diff}
\end{figure}

Fig.~\ref{fig:Pk-diff}
shows the difference of the power spectrum
between the case of 3LPT initial conditions and
other initial conditions. 
The difference of the
power spectrum between 3LPT and 1LPT is several percents during
$0 < z < 10$. The difference becomes maximum
at $z \simeq 3$. The difference of the power
spectrum between the case of ZA initial conditions
and higher-order initial conditions
becomes about several percents in small scale.
Especially high precision is required for the power
spectrum such a case of BAO analysis,
at least 2LPT initial condition should be 
considered. 
In other words, when we require $1 \%$ accuracy
for the power spectrum,
we must generate the initial conditions for the cosmological
simulations at least 2LPT. Especially when we consider
deep survey ($z >1$), the choice of the initial conditions
will become still more important. 
In near future, the precision
cosmology would require sub percents accuracy. In this case,
we will generate the initial conditions for the cosmological
simulations at least 3LPT. 
Even if sub percents accuracy is required, the transverse
mode in 3LPT is neglegible.

\subsection{Non-Gaussianity} \label{sec:NonGauss}

Next, we notice a a one-point probability distribution function
of the density fluctuation field ${\mathcal P} (\delta)$
the probability of obtaining the value $\delta$
plays an important role.
If $\delta$ is a random Gaussian field, 
the PDF of the density 
fluctuation is determined as
\begin{eqnarray}
{\mathcal P}(\delta) = \frac{1}{(2\pi \sigma^2)^{1/2}}
 e^{-\delta^2/2\sigma^2}\,,
\label{Gauss}
\end{eqnarray}
where 
$\sigma^2 \equiv \left < \left (\delta-\left <\delta 
\right> \right )^2 \right > $ 
is the dispersion and $\left <\;\;\right >$ denotes 
the spatial average.

In this paper, we set the initial condition by random Gaussian
field. If the growth of the density fluctuation is linear,
the density field still be Gaussian. 
The expected tools to detect
transients at large scales are 
the cumulants of the one-point PDF of the density fluctuation
field whose orders are higher than two
which  become nonzero for the distribution
deviating from Gaussian.
In the structure formation, non-Gaussianity are 
generated because of the non-linear dynamics
of the fluctuations, even though $\delta$
is initially treated as a random Gaussian field, 
as a result of the generic prediction of inflationary
scenario.

For this purpose, we concentrate on
the third and fourth order cumulants,
which are defined as 
$<\delta^3>_c \equiv <\delta^3>$,
$<\delta^4>_c \equiv <\delta^4>-3\sigma^4$
display asymmetry and non-Gaussian degree of
``peakiness'', respectively, for a given dispersion
\cite{Peebles,Peacock}. Since
it is known that the scaling 
$<\delta^n>_c \propto \sigma^{2n-2}$ holds
for weakly non-linear regions during 
the gravitational clustering 
\cite{Bernardeau2002}
from Gaussian initial conditions,
 we introduce 
the following normalized higher-order statistical
quantities~:
\begin{eqnarray*}
\mbox{skewness} &:& \gamma = \frac{ \left < \delta ^3 \right >_c }
{\sigma^4} \,, \\
\mbox{kurtosis} &:& \eta = 
\frac{\left < \delta^4 \right >_c}{\sigma^6} \,.
\end{eqnarray*}
The merit of adopting these definitions is, as stated above,
that they are constants in weakly nonlinear stage
which are given by Eulerian linear and second-order
perturbation theory \cite{Peebles,Bernardeau2002}. 
For example, 
in the E-dS model smoothed with a spherical top-hat
window function (Eq.~(\ref{tophat_window})),
the skewness and the kurtosis are given by
\begin{eqnarray}
\gamma &=& \frac{34}{7} + y_1 + {\cal{O}}(\sigma^2) \, ,\\
\eta &=& \frac{60712}{1323} + \frac{62}{3}y_1
-\frac{7}{3}y_1^3 + \frac{2}{3}y_2 
+{\cal{O}}(\sigma^2) \, ,
\end{eqnarray}
where
\begin{eqnarray}
y_p \equiv  \frac{d^p \ln \sigma^2(R)}
{d \ln^p R},
\end{eqnarray}
with smoothing scale $R$~\cite{Bernardeau2002}.

For this form of the skewness and the kurtosis,
the effects of transients at large scales
from the ZA initial condition is also investigated by
\cite{Scoccimarro:1997gr,Valageas:2001} 
as 
\begin{eqnarray}
\gamma_{\rm tran} &=& - \frac{6}{5a} + \frac{12}{35 a^{7/2}}\,,
\label{transients_ZA_initial_skew} \\
\eta_{\rm tran} &=& -\frac{816}{35a}-\frac{28 y_1}{5a}
+\frac{184}{75a^2}+\frac{1312}{245a^{7/2}}
\nonumber\\
&&+\frac{8 y_1}{5 a^{7/2}}-\frac{1504}{4725a^{9/2}}
+\frac{192}{1225a^7}\,.
\label{transients_ZA_initial_kurt}
\end{eqnarray}

In the analysis of the non-Gaussianity, as we showed
in Paper I, the dispersion among samples
is quite large. In this subsection, we do not show 
error bars in figures.

\begin{figure}[tb]
\centerline{
\includegraphics[height=13cm]{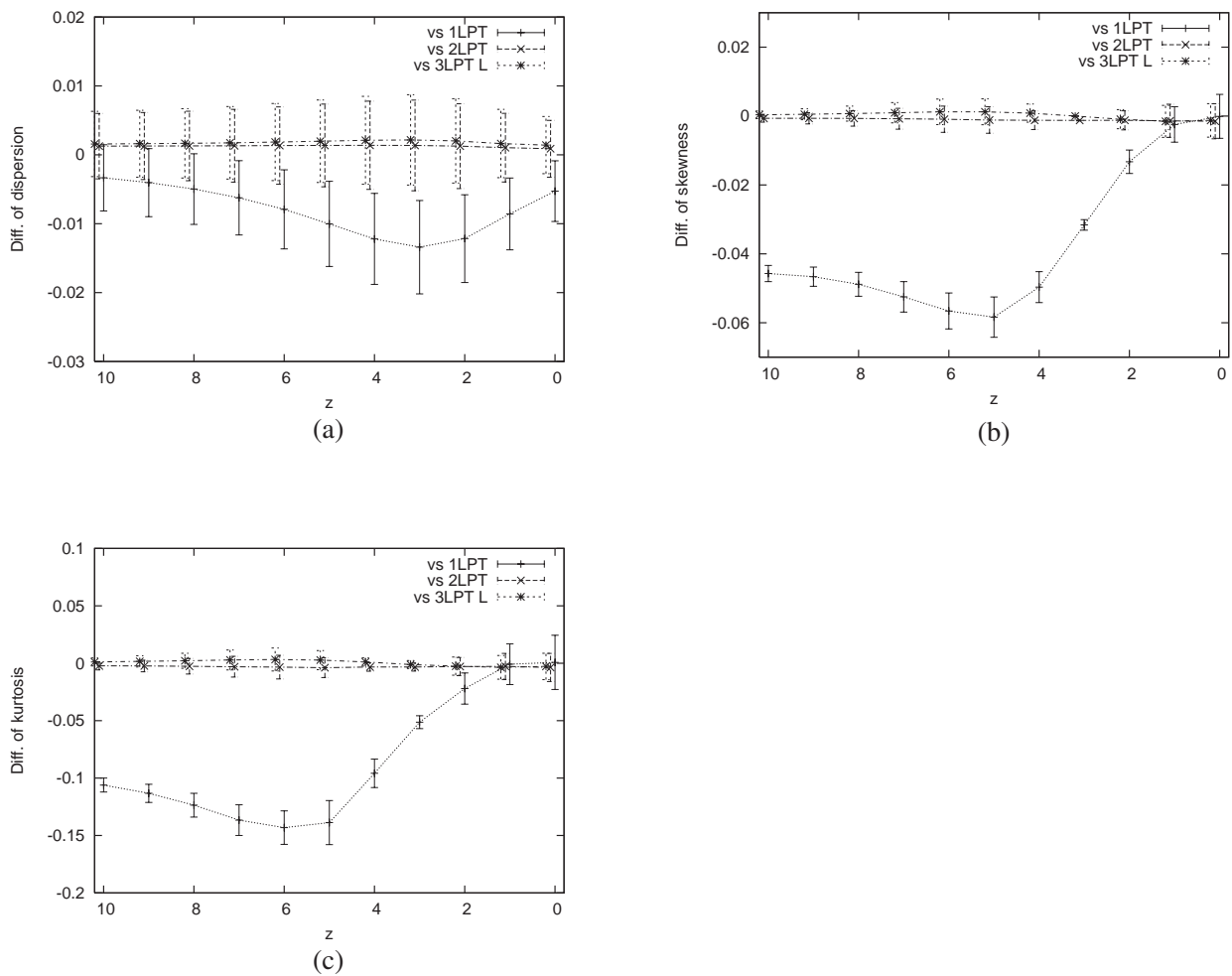}
}
\caption{The difference of the dispersion and non-Gaussianity
of the density distribution
from N-body simulation ($R = 1 h^{-1}$ Mpc).
Here we compare 3LPT and other perturbations. 
In order to make the difference legible, plots are slightly shifted. 
(a) The difference of the density dispersion. 
(b) The difference of the skewness.
(c) The difference of the kurtosis.}
\label{fig:Diff-NG}
\end{figure}

Fig.~\ref{fig:Diff-NG} shows the difference of 
the evolution of the density dispersion and non-Gaussianity.
Here we choose smoothing scale $R=1 h^{-1}$ [Mpc]. 
The dispersion increases monotonically. 
The difference of the dispersion 
with the initial conditions based on 3LPT and ZA
is less than $2 \%$. In other words,
the effect of higher-order perturbations seems percent
order. Then the difference between dispersion 
with the initial conditions based on the 2LPT and 3LPT
perturbations is less than $0.1 \%$. For the dispersion,
the effect of 3LPT seems almost neglectable.

The difference between non-Gaussianity with the initial 
conditions based on the ZA and higher-order
perturbations
seems percent order. Therefore, when we require
$1 \%$ accuracy for the non-Gaussianity, at least
we should apply 2LPT for the initial conditions.
Furthermore, when we require
$0.1 \%$ accuracy for the non-Gaussianity, at least
we should apply 3LPT for the initial conditions.
For non-Gaussianity, the effect of 3LPT transverse
mode is quite small.
Because the difference of non-Gaussianity
with the initial conditions based on 3LPT varies
between samples, it is hard to discuss the effect of
the transverse mode in 3LPT for the initial condition.

\section{Summary} \label{sec:Summary}

Recently, observations constrained several cosmological constants.
In order to compare theoretical predictions with observations,
high accuracy for cosmological simulations has been required.
In precise simulations, a setup of initial conditions becomes
one of most important problem. 

In this paper, we have discussed initial condition problem
for cosmological $N$-body simulations with Lagrangian
perturbation theory (LPT). In standard method, Zel'dovich
approximation (ZA), i.e., first-order LPT (1LPT) have been
applied for the initial conditions for a long time.
However, it is well known that ZA is insufficient for description
of the evolution for non-Gaussianity in the density field. 
The reason is the existence of so-called transients, that is, the most
dominant decaying mode arising from our ignorance of the
initial conditions.
Crocce \textit{et al.}~\cite{Crocce2006} pointed out the effect of the
transient for non-Gaussianity with initial conditions based on
ZA and 2LPT. They show that although 2LPT decreases the effect 
of the transient compared with ZA, it still remains. 
In statistics quantities, such as non-Gaussianity and the spectrum, 
the difference of several percent appears among both. 
Then in Paper I~\cite{Tatekawa2007},
we analyzed the non-Gaussianity with the 
initial conditions based on 1LPT, 2LPT, and 3LPT.
Since 3LPT initial conditions are expected to provide exact
results for the kurtosis in the weakly non-linear region,
we can evaluate the impact of transients from 2LPT initial conditions. 
The initial conditions are set up at $z \simeq 20, 30, 80$.
When we set the initial conditions at $z \simeq 80$, the
difference between the non-Gaussianity with the initial conditions
based on 2LPT and 3LPT almost disappears. In Paper I,
the transverse mode in 3LPT was ignored.

We have analyzed the effect of the transverse mode
in 3LPT. 
In addition to non-Gaussianity,
the power spectrum has been
analyzed. In all the comparison, the difference of statistical
quantities with the initial conditions based on 2LPT and 3LPT is
sub-percent order. In other words, when we require more than
$1$ \% accuracy, we should set up initial conditions for
cosmological $N$-body simulations with 3LPT. Then the
effect of the transverse mode in 3LPT seems quite small.
In the analysis for density field, the transverse mode in 3LPT
would be ignored. 

Why the impact of the transverse mode in 3LPT is quite small
for the density field? We can explain the reason by the 
description of the density fluctuation with the Lagrangian
perturbation. Here we expand Eq.(\ref{eqn:Jacobian}).
\begin{eqnarray} \label{eqn:Jacobian-exp}
\delta &=& - \nabla \cdot \bm{s} + \left ( \nabla \cdot \bm{s} \right )^2
 -\frac{1}{2} \left ( \left ( \nabla \cdot \bm {s} \right )^2
 - s_{i,j} s_{i,j} \right )  \nonumber \\
  &&
 -  \left ( \nabla \cdot \bm{s} \right )^3
 + \left ( \nabla \cdot \bm{s} \right )
  \left ( \left ( \nabla \cdot \bm {s} \right )^2
 - s_{i,j} s_{i,j} \right )  \nonumber \\
 &&
 - \det \left ( s_{i,j} \right ) + \cdots  \,.
\end{eqnarray}
The leading order of Eq.(\ref{eqn:Jacobian-exp})
is given by the divergence of the Lagrangian displacement vector.
The longitudinal mode of 3LPT causes third-order effect
for the density fluctuation. On the other hand, by the definition,
the divergence of the transverse mode of 3LPT disappears.
Therefore the transverse mode of 3LPT gives 
fourth-order effect for the density fluctuation. Here we set up
the initial conditions in high-$z$ region,
the impact of the transverse mode would be quite small.
When we have interested in peculiar velocity
or the density field in redshift space, the transverse mode in 3LPT
would affect as large as the longitudinal mode in 3LPT. 

Finally, we briefly mention applications of our study to
observations of the near future. Several projects which survey
galaxies around $z=1$ are carrying out or
planning~\cite{BOSS,DEEP2}. 
These results
show not only the result of structure formation but also its evolution.
By comparison of theoretical prediction and observation,
it is expected that severe restriction is given to the nature of dark energy. 
Quite high-precision theoretical prediction is required for this comparison. 
Our study about the initial conditions will contribute to cosmological
simulations for restricting dark energy models.

\ack
We thank Nobuyoshi Komatsu, Takahiko Matsubara,
and Shuntaro Mizuno for useful discussions.
GRAFIC code was carried out on computer systems
in astrophysics group, Ochanomizu University.

\end{document}